\documentclass[runningheads]{llncs}
\usepackage{graphicx}
\usepackage{float}
\usepackage{subfigure}
\usepackage{epsfig}
\usepackage{color}
\usepackage{amsmath}
\begin{document}
\title{Fully Automated Left Atrium Cavity Segmentation from 3D GE-MRI by Multi-Atlas Selection and Registration}
%
%

\author{Mengyun Qiao\inst{1} \and
Yuanyuan Wang\inst{1} \and
Rob J. van der Geest\inst{2} \and Qian Tao \inst{2}}
\authorrunning{Qiao et al.}
%
\institute{Biomedical Engineering Center, Fudan University, Shanghai, China \and
Department of Radiology, Leiden University Medical Center, Leiden, the Netherlands}
%
\maketitle              
\begin{abstract}
This paper presents a fully automated method to segment the complex left atrial (LA) cavity, from 3D Gadolinium-enhanced magnetic resonance imaging (GE-MRI) scans. The proposed method consists of four steps: (1) preprocessing to convert the original GE-MRI to a probability map, (2) atlas selection to match the atlases to the target image, (3) multi-atlas registration and fusion, and (4) level-set refinement. The method was evaluated on the datasets provided by the MICCAI 2018 STACOM Challenge with 100 dataset for training. Compared to manual annotation, the proposed method achieved an average Dice overlap index of 0.88.

\end{abstract}
%
%
%
\section{Introduction}
The left atrial (LA) cavity segmentation is an important step in reconstructing and visualizing the patient-specific atrial structure for clinical use. Various important clinical parameters can be derived from accurately segmented LA structures\cite{haissaguerre1998spontaneous,AF_ablation}. With development of imaging techniques, the three-dimensional LA geometry can be non-invasively visualized by computed tomography angiography (CTA) or magnetic resonance angiography (MRA). Gadolinium-enhanced magnetic resonance imaging (GE-MRI) is typically used to study the extent of fibrosis (scars) across the LA, which provides clinically important diagnostic and prognostic information. However, manual segmentation of the atrial chambers from GE-MRI is a highly challenging and time-consuming task, due to the complex shape of LA, as well as the low contrast between the atrial tissue and background. Accurate automated computer methods to segment the LA and reconstruct it in three dimensions are highly desirable\cite{mcgann2014atrial,hansen2015atrial,zhao2017three}.

In this work, we present and evaluate a systematic workflow to segment the LA in a fully automated manner. The proposed workflow consists of four steps: (1) preprocessing to convert the original GE-MRI to a probability map, (2) atlas selection, (3) multi-atlas registration and fusion, and (4) level-set refinement.

\section{Data and Method}\label{section2}
\subsection{Data}\label{subsection1}
The majority of the dataset used in this work were provided by The University of Utah (NIH/NIGMS Center for Integrative Biomedical Computing (CIBC)), and the rest were from multiple other centers. For all clinical data, institutional ethics approval was obtained. 

In total, 100 datasets were provided with annotated reference standard for training purposes.Each 3D MRI patient data was acquired using a clinical whole-body MRI scanner and contained the raw MRI scan and the corresponding ground truth labels for the LA cavity and pulmonary veins. The ground truth segmentation was performed by expert observers.  

\subsection{Probability Map}\label{subsection1}
\begin{figure}[!htbp]
	\centering 	
	\includegraphics[width=5in]{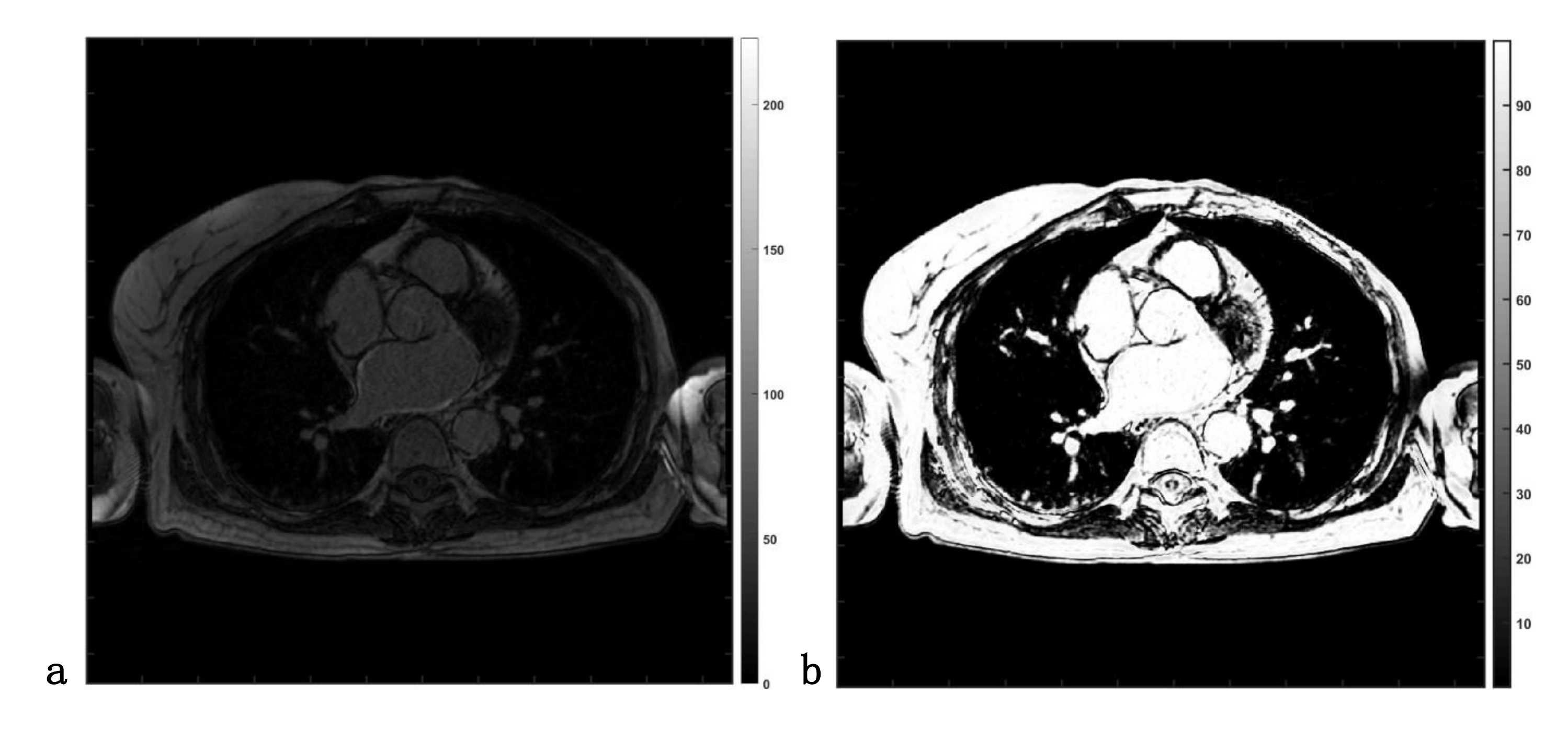} 
	\caption{An example of the GE-MRI slice: (a) the original GE-MRI scan (arbitrary range), (b) the preprocessed probability map (range 0-100$\%$).} 
	\label{Fig_map}
\end{figure}

GE-MRI typically has non-quantitative signal intensity values that vary among different scans. The difference is especially pronounced if the scans are from different vendors or centers. To normalize the signal intensity value, we converted the original GE-MRI to a probability tissue map as a preprocessing step, with the signal intensity value normalized between 0 and 100$\%$. This was realized by the Coherent Local Intensity Clustering (CLIC) algorithm \cite{li2009mri}: the class membership functions were calculated by a non-supervised clustering algorithm, and the probability maps were derived accordingly. The preprocessing was applied on the entire 3D volume of GE-MRI. Fig. \ref{Fig_map} shows a slice of the original GE-MRI and the resulting probabilistic image.

{\color{red} }

\subsection{SIMPLE Atlas Selection}\label{subsection2}
Atlas images were selected based on image similarity between the atlas and the target. We used the selective and iterative method for performance level estimation (SIMPLE) method for atlas selection based on 3D atrial anatomy \cite{langerak2010label}. Poorly performing atlases can be regarded as noise in the majority voting. These segmentations can be down-weighted by their estimated performance to reduce their influence on the result of the majority voting. We used the groupwise registration to get the propagated segmentations from atlas $L$, and then the segmentations were combined into $L_{mean}$ using a weighted majority vote. $L_{mean}$ was assumed to be reasonably similar to $L_{target}$ \cite{artaechevarria2009combination}. The estimated performance was recomputed as $f(L_{i},L_{mean}^{0})$, where $f$ is the Dice overlapping index, $i$ is the atlas index, and 0 denotes the number of iterations. The poorly performing atlas images were discarded if is lower than a predefined threshold, while the rest was combined into a new atlas set for next iteration of groupwise registration. The iteration was stopped when the atlas set converged.

\subsection{Multi-Atlas Segmentation by Groupwise Registration}\label{subsection3}
Mutli-atlas segmentation is an image-based approach to segment complex anatomical structures in medical images \cite{atlas_survey}. Define $A_i$ as the atlas image, with known label $L_i$, $i=1,2,...,N$, where $N$ is the total number of atlases. Multiple images with labels were used as a base of knowledge, to segment a new image $I$ containing the same anatomical structure. 

In this work, we used the groupwise strategy to perform multi-atlas registration \cite{Yigitsoy_groupwise2011,Wachinger2013}. We registered all the atlas images in one groupwise registration to the given image. We formulated it as optimizing a group objective function that merges the given image with the atlas images:

\begin{equation}
{\hat{\boldsymbol{\mu}}}=\arg \underset{\boldsymbol{\mu}}{\min}\boldsymbol{C}({\boldsymbol{T_{\mu}} ;I,A_1,A_2,...,A_N)}
\label{eq_groupatlas}
\end{equation}
where the cost function $C$ is defined to minimize the variance in the group of images:
\begin{equation}
\boldsymbol{C}({\boldsymbol{\mu}})= (\boldsymbol{T_{\mu}}(I)-\bar{I}_{\boldsymbol{\mu}})^2 + \sum_{i=1}^N(\boldsymbol{T_{\mu}}(A_i)-\bar{I}_{\boldsymbol{\mu}})^2
\label{eq4}
\end{equation}
with $\bar{I}_{\boldsymbol{\mu}}$ is defined as:
\begin{equation}
\bar{I}_{\boldsymbol{\mu}}=\frac{1}{N+1} \left\{ \boldsymbol{T_{\mu}}(I)+\sum_{i=1}^{N} \boldsymbol{T_{\mu}}(A_i)  \right\}
\label{eq5}
\end{equation}
The transformation parameter $\boldsymbol{\mu}$ is the ensemble of all transformation applied to $N$ atlas images $A_i$ and the given image $I$. Given the preprocessing step, the signal intensity range in the atlas images and target image is comparable. Therefore minimization of the variance indicates that all images in the groupwise registration are well registered. The propagated labels were fused by the majority-vote method to obtain the segmentation result.

\subsection{Refinement of LA Segmentation by Level-Set}\label{subsection4}
\begin{figure}[!htbp]
	\centering 	
	\includegraphics[width=5in]{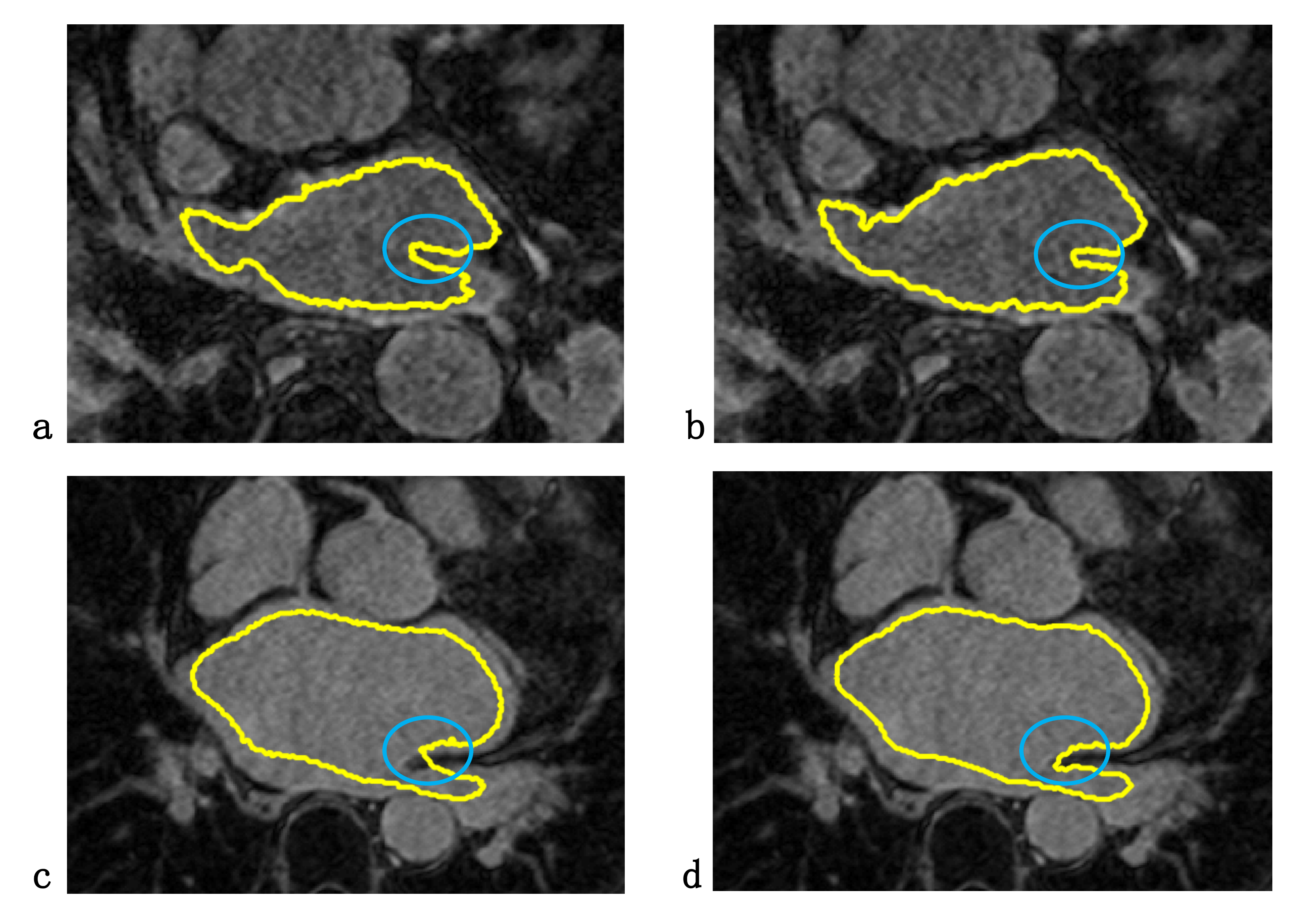} 
	\caption{Two examples of the level-set refinement: (a) and (c) are before refinement with dice index of  0.89, while (b) and (d) are after refinement with dice index of  0.91. The blue circle highlights the area of improvement.} 
	\label{Fig_levelset}
\end{figure}

From the atlas-based segmentation, the local methods can still provide incremental improvement. In this step, we used a level-set approach to refine and grow the initial segmentation, attending to local image details. With the atlas-based segmentation as initialization, the level-set \cite{Chan2001Active} was applied with the energy function formulated as:
\begin{equation}
\begin{aligned} 
F(\phi )=\mu\int_{\Omega }\left | \nabla H(\phi ) \right |\mathrm{d}\Omega +\nu \int_{\Omega } H(\phi)\mathrm{d}\Omega\\
+\lambda _1\int_{\Omega}\left | I(x,y,z)-c_1 \right |^2H(\phi )\mathrm{d}\Omega\\
+\lambda _2\int_{\Omega}\left | I(x,y,z)-c_2 \right |^2(1-H(\phi ))\mathrm{d}\Omega
\label{eq6}
\end{aligned}
\end{equation}
where $\phi$ is the level set function. The model assumes that the image $I$ is a three-dimensional image with piecewise constant values. It defines the evolving surface $\Omega=0$ as the boundary of object to be detected in image $I$. $c_1$, $c_2$ are the average values inside and outside the boundary, respectively. $H$ is the Heaviside function. $ \mu \ge 0 $, $ \nu \ge 0 $, $\lambda _1,\lambda _2> 0$ are weighting factors of the four terms. The first term denotes the surface area, and second term denotes the volume inside the surface. The last two terms are the variance inside and outside the boundary. A characteristic of the level-set method approach is that it does not have prior assumptions on the object geometry and can deal with complex shapes such as the LA.  Fig. \ref{Fig_levelset} shows an example of the refinement step.

\section{Experiments and Results}\label{section3}
\begin{figure}[!htbp]
	\centering 	
	\includegraphics[width=5in]{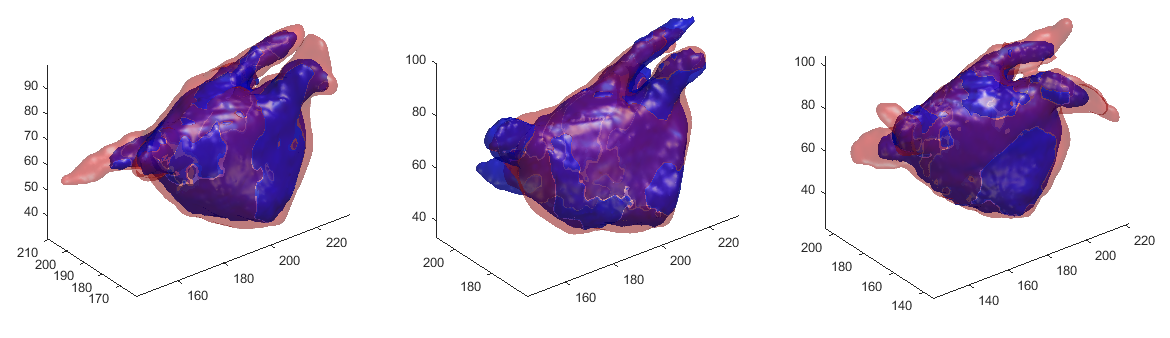} 
	\caption{3D reconstruction of three examples from the provided datasets: red is the manual ground truth segmentation, and blue is the automated segmentation.} 
	\label{Fig_result}
\end{figure}

Our proposed method was evaluated on the training datasets (N=100)  and the testing datasets(N=54) datasets provided by the MICCAI 2018 STACOM challenge organizers. The leave-one-out strategy was used: for one target image, we used the rest of 99 datasets as the potential atlases. We carried out preliminary experiments and compared the performance of two atlas selection methods: (1) random atlas selection, (2) SIMPLE atlas selection. In each method 10 best atlases were finally selected. The performance in terms of Dice index is reported in Table.\ref{Table1}. The proposed method that tested on the training datasets resulted in an average Dice index of $0.88 \pm 0.03$ and average perpendicular distance (APD) of $2.47 \pm 1.01$ mm, superior to random atlas selection (P$<$0.05 by the Paired Wilcoxon-test). The method tested on the testing datasets resulted in an average Dice index of $0.93$ . Fig. \ref{Fig_result} shows three examples of the 3D reconstruction of the LA cavity segmentation by the proposed method, compared to the ground truth manual segmentation.

\begin{table}[]
\centering
\caption{ Performance of our proposed method on the training datasets in terms of average perpendicular distance(APD) and Dice index. The Dice and APD after applying the level-set refinement are also reported.}
\label{Table1}
\begin{tabular}{llll}
\hline
\multicolumn{4}{l}{Evaluation after atlas segmentation}     \\ \hline
             & Random-selection & SIMPLE-selection & P-value \\
Dice indices & $0.83\pm 0.07$  & $0.87\pm 0.04$   & $<0.05$  \\
APD          & $2.67\pm 0.93$  & $2.49\pm 0.01$   & $<0.001$ \\ \hline
\multicolumn{4}{l}{Evaluation after level-set refinement}   \\ \hline
             & Random-selection & SIMPLE-selection & P-value \\
Dice indices & $0.84\pm 0.08$  & $0.88\pm 0.03$   & $<0.05$  \\
APD          & $2.65\pm 0.84$  & $2.47\pm 1.01$   & $<0.001$ \\ \hline
\end{tabular}
\end{table}

\section{Discussion and Conclusion}\label{section4}
In this work, we have developed an atlas-based method to segment the complex LA cavity from GE-MRI scans. The method includes atlas selection, registration, and label fusion. In addition, for better performance of the atlas approach, we have normalized the GE-MRI images by the probability map, and refined the final results using the level-set approach.

Depending on similarity between the given image to the individual atlases, we used iterative rules to select suitable atlases. The atlas-selection method estimated the performance of segmentation of an individual atlas in an iterative manner, such that the best performed atlases were selected in \emph{ a posteriori} manner for progressively better segmentation performance.

We believe that the atlas approach is still interesting in face of the current trend of deep-learning-based medical image segmentation. Especially when the training data is scarce, the atlas approach preserves the geometrical characteristics of anatomical structures; by comparing a target image with the atlases, the algorithm still performs a type of ``learning", although not data-intensive. 
 
In conclusion, we have presented a complete workflow to fully automatically segment the LA cavity from GE-MRI. Trained and tested on the MICCAI 2018 STACOM Challenge database, the proposed method showed reasonably high accuracy of LA segmentation compared to the time-consuming manual annotation.

%
%
%
%
%
\section*{Acknowledgements}
This research is supported by grants from National Key R$\&$D Program of China 2018YFC0116303.
\bibliography{mybibfile}

\begin{thebibliography}{10}
\providecommand{\url}[1]{\texttt{#1}}
\providecommand{\urlprefix}{URL }
\providecommand{\doi}[1]{https://doi.org/#1}

\bibitem{artaechevarria2009combination}
Artaechevarria, X., Munoz-Barrutia, A., Ortiz-de Sol{\'o}rzano, C.: Combination
  strategies in multi-atlas image segmentation: application to brain mr data.
  IEEE transactions on medical imaging  \textbf{28}(8),  1266--1277 (2009)

\bibitem{Chan2001Active}
Chan, T.F., Vese, L.A.: Active contours without edges. IEEE Transactions on
  Image Processing A Publication of the IEEE Signal Processing Society
  \textbf{10}(2), ~266 (2001)

\bibitem{haissaguerre1998spontaneous}
Haissaguerre, M., Ja{\"\i}s, P., Shah, D.C., Takahashi, A., Hocini, M.,
  Quiniou, G., Garrigue, S., Le~Mouroux, A., Le~M{\'e}tayer, P., Cl{\'e}menty,
  J.: Spontaneous initiation of atrial fibrillation by ectopic beats
  originating in the pulmonary veins. New England Journal of Medicine
  \textbf{339}(10),  659--666 (1998)

\bibitem{hansen2015atrial}
Hansen, B.J., Zhao, J., Csepe, T.A., Moore, B.T., Li, N., Jayne, L.A.,
  Kalyanasundaram, A., Lim, P., Bratasz, A., Powell, K.A., et~al.: Atrial
  fibrillation driven by micro-anatomic intramural re-entry revealed by
  simultaneous sub-epicardial and sub-endocardial optical mapping in explanted
  human hearts. European heart journal  \textbf{36}(35),  2390--2401 (2015)

\bibitem{atlas_survey}
Iglesiasa, J.E., Sabuncu, M.R.: Multi-atlas segmentation of biomedical images:
  A survey. Medical Image Analysis  \textbf{24}(1),  205--219 (2015)

\bibitem{langerak2010label}
Langerak, T.R., van~der Heide, U.A., Kotte, A.N., Viergever, M.A., Van~Vulpen,
  M., Pluim, J.P.: Label fusion in atlas-based segmentation using a selective
  and iterative method for performance level estimation (simple). IEEE
  transactions on medical imaging  \textbf{29}(12),  2000--2008 (2010)

\bibitem{li2009mri}
Li, C., Xu, C., Anderson, A.W., Gore, J.C.: Mri tissue classification and bias
  field estimation based on coherent local intensity clustering: A unified
  energy minimization framework. In: International conference on information
  processing in medical imaging. pp. 288--299. Springer (2009)

\bibitem{mcgann2014atrial}
McGann, C., Akoum, N., Patel, A., Kholmovski, E., Revelo, P., Damal, K.,
  Wilson, B., Cates, J., Harrison, A., Ranjan, R., et~al.: Atrial fibrillation
  ablation outcome is predicted by left atrial remodeling on mri. Circulation:
  Arrhythmia and Electrophysiology  \textbf{7}(1),  23--30 (2014)

\bibitem{AF_ablation}
Oral, H., Pappone, C., Chugh, A., Good, E., Bogun, F., Pelosi, F., Bates, E.R.,
  Lehmann, M.H., Vicedomini, G., Augello, G., Agricola, E., Sala, S.,
  Santinelli, V., Morady, F.: Circumferential pulmonary-vein ablation for
  chronic atrial fibrillation. N Engl J Med  \textbf{354},  934--941 (2006)

\bibitem{Wachinger2013}
Wachinger, C., Navab, N.: Simultaneous registration of multiple images:
  Similarity metrics and efficient optimization. IEEE Transactions on Pattern
  Analysis and Machine Intelligence  \textbf{35}(5),  1221--1233 (May 2013).
  \doi{10.1109/TPAMI.2012.196}

\bibitem{Yigitsoy_groupwise2011}
Yigitsoy, M., Wachinger, C., Navab, N.: Temporal groupwise registration for
  motion modeling. In: Proceedings of the 22Nd International Conference on
  Information Processing in Medical Imaging. pp. 648--659. IPMI'11,
  Springer-Verlag, Berlin, Heidelberg (2011),
  \url{http://dl.acm.org/citation.cfm/id=2029686.2029745}

\bibitem{zhao2017three}
Zhao, J., Hansen, B.J., Wang, Y., Csepe, T.A., Sul, L.V., Tang, A., Yuan, Y.,
  Li, N., Bratasz, A., Powell, K.A., et~al.: Three-dimensional integrated
  functional, structural, and computational mapping to define the structural
  ¡°fingerprints¡± of heart-specific atrial fibrillation drivers in human heart
  ex vivo. Journal of the American Heart Association  \textbf{6}(8),  e005922
  (2017)

\end{thebibliography}

\end{document}